# EXTENDING THE ER MODEL TO RELATIONAL MODEL NOVEL TRANSFORMATION ALGORITHM: TRANSFORMING RELATIONSHIP TYPES AMONG SUBTYPES

Dhammik Pieris, dhammika.pieris@monash.edu

Monash University

## *ABSTRACT*

A novel approach for creating ER conceptual models and an algorithm for transforming them to the relational model has been developed by modifying and extending the existing methods. A part of the new algorithm has previously been presented. This paper presents the rest of the algorithm. One of the objectives of this paper is to use it as a supportive document for ongoing empirical evaluations of the new approach being conducted using the cognitive engagement method and with the participation of different segments of the field as respondents.

## 1. INTRODUCTION

A novel approach for creating ER conceptual models(Pieris, 2013a) and an algorithm(Pieris, 2013b) for transforming them to the relational model has been developed by modifying and extending the existing methods. A part of the new algorithm has previously been presented(Pieris, 2013b). This paper presents the rest of the algorithm. The new algorithm contains 14 steps, out of which 6 steps were published in a previous occasion. An early version of the algorithm and the reasons for the need of a novel approach can be found in (Pieris & Rajapakse, 2012).

A summary of the remaining 8 steps are given in the table below. Each of the step and how it can be applied will be demonstrated using different mini-ER models. For example the first step: Step GOG is demonstrated with the help of an ER model given in the Figure 1, etc.

| Table 1: Summary of the Steps of the remaining Algorithm | | |
|---|---|---|
| No | Step | What to transform |
| 1 | GOG | One-to-one (1:1) relationship type between two regular entity type |
| 2 | GMG | Many-to-many (M:N) relationship type between two regular entity type |
| 3 | SNG | One-to-many (1:N) relationship type between a subtype and a regular entity type |
| 4 | SMG | M:N relationship type between a subtype and a regular entity type |
| 5 | SOS | One-to-one relationship type between two subtypes |
| 6 | SNS | One-to-ma**n**y relationship type between two **s**ubtypes |
| 7 | SMS | Many-to-many relationship type between two subtypes |
| 8 | THG | Ternary and higher order relationship types among regular entity types |

**Step GOG: Transforming a 1:1 relationship type between two regular entity types**

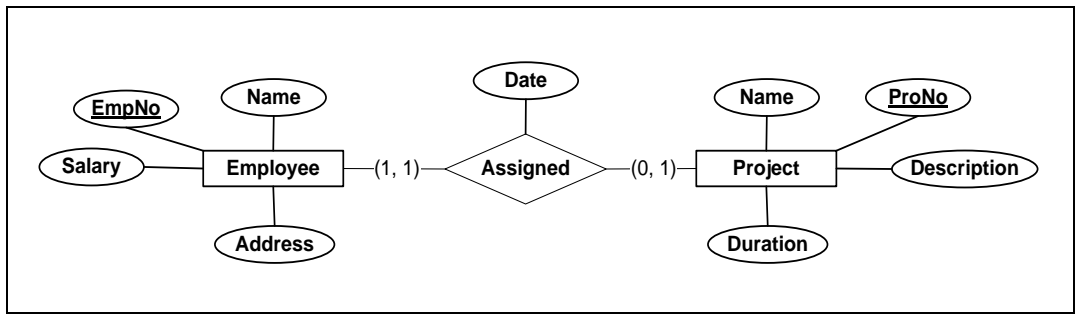

**Figure 1. An ER Model with a 1:1 relationship Type between two regular Entity Types**

Follow the steps below to transform a 1:1 relationship type that can be existed between two regular entity types.

*1. Chose one of the two regular entity type relations–"Employee", say. It is recommended choosing a relation analogy to an entity type with "total participation"[1] in the relationship type. Include as a foreign key (FK) in the chosen relation following its last attribute, the PK of the non-chosen relation,*

   Employee[EmpNo, Name, Address, Salary, ProNo]

*2. To the including FK, give a bracketed suffix of 4 variables and set the name of the relationship type as the 1st variable.*

   Employee[EmpNo, Name, Address, Salary, ProNo(Assigned, , , )]

*3. Chose the remaining 3 variables from the values of the cardinality ratio pairs associated with the relationship type as follows*

   (i) *As the 2nd variable, chose the min value of the cardinality ratio pair that lies nearest to the entity type that corresponds to the chosen relation. Note that the max value of the same cardinality ratio pair would not be transformed and assumed to be 1 always.*

   Employee[EmpNo, Name, Address, Salary, ProNo(Assigned, 1 , , )]

   (ii) *As the 3rd and 4th variables choose the min & max values of the cardinality ratio pair that lies farthest from the entity type that corresponds to the chosen relation,*

   Employee[EmpNo, Name, Address, Salary, ProNo(Assigned, 1 , 0 , 1) ]

*4. Include any set of simple attributes, if existed, of the relationship type as attributes of the chosen relation following the FK included*

---

[1] If the min value of a cardinality ratio pair corresponding to a *participation* of an entity type in a relationship type is non-zero, then the *participation* is said to be a "*total participation*".

Employee[EmpNo, Name, Address, Salary, ProNo(Assigned, 1 , 0 , 1), Date ]

Thus, the final RDS of the ER model in Figure 1 is as follows.

Project[ProNo, Name, Description, Duration]
Employee[EmpNo, Name, Address, Salary, ProNo(Assigned, 1 , 0 , 1), Date ]

**Step GMG: Transforming binary M:N relationship type between two regular entity types**

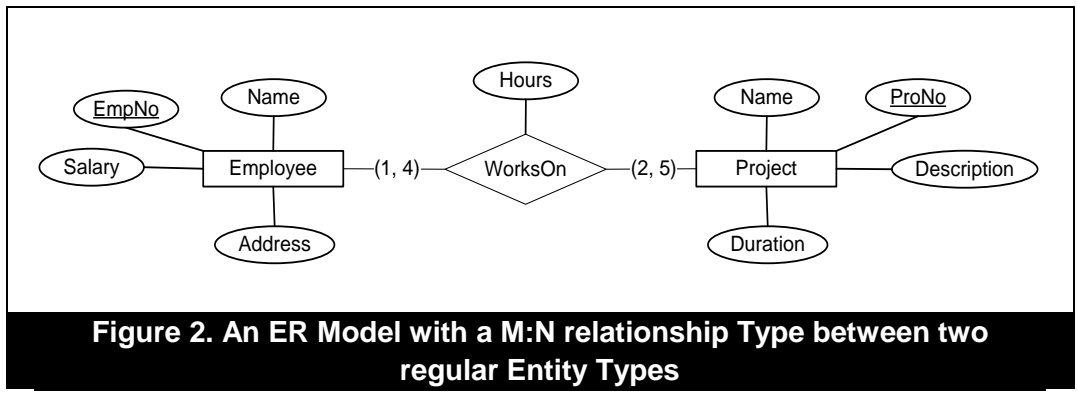

**Figure 2. An ER Model with a M:N relationship Type between two regular Entity Types**

Follow the steps below to transform a M:N relationship type that can be existed between two regular entity types.

1. *Create a new relation by the name of the relationship type*,

    WorkOn[ ]

2. *As the first two attributes of the new relation, include PKs: of the relations corresponding to the participative entity types and underline them separately. Both PKs will act together as the PK of the new relation, and each of them acts individually as an FK referring to it's corresponding relation*

WorkOn[EmpNo, ProNo]

3. *Give a bracketed suffix of 2 variables to each of the including PK, and include as the two variables, the pair of min-max-cardinality-ratio-values corresponding to the participation in the relationship type of the entity type analogy to the PK considered.*

WorkOn[EmpNo(1,4), ProNo(2,5)]

4. *Include any attribute attached to the relationship type as the remaining attributes of the new relation following both of the suffixed FKs in the new relation.*

WorkOn[EmpNo(1, 4), ProNo(2, 5), Hours]

With this transformation the RDS of the ER model in the Figure 1 will be changed to the following.

Employee[EmpNo, Name, Address, Salary]
Project[ProNo, Name, Description, Duration]
WorkOn[EmpNo(1, 4), ProNo(2, 5), Hours]

**Step SNG: Transforming a binary 1:N relationship type between a subtype and a regular entity type**

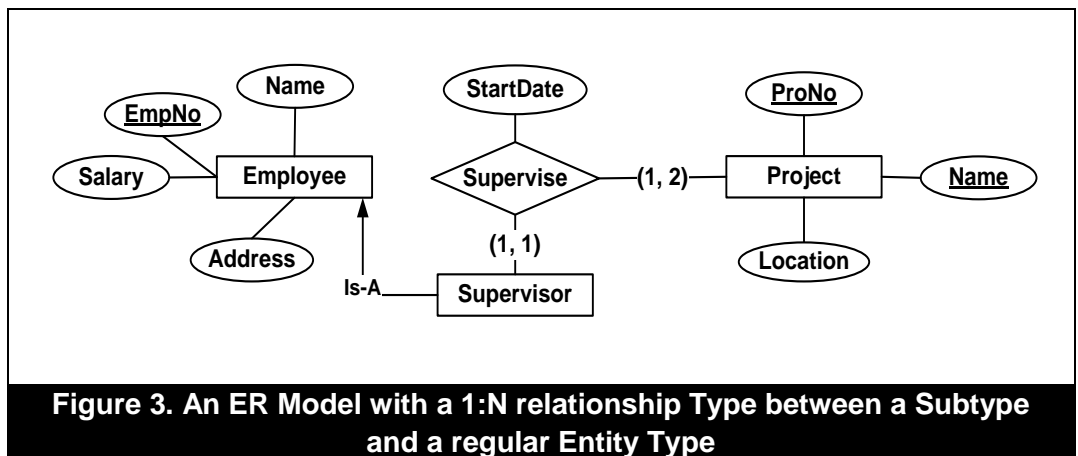

**Figure 3. An ER Model with a 1:N relationship Type between a Subtype and a regular Entity Type**

Follow the steps below to transform a 1:N relationship type that can be existed between a subtype and a regular entity type.

1. *Chose the relation analogy to the entity type lies at the N-side of the relationship type.*

    Supervisor[EmpNo]

2. *Include the PK analogy to the non-chosen relation, as an FK in the chosen relation following its last attribute*

    Supervisor[EmpNo, ProNo]

3. *To the including FK, give a bracketed suffix of 4 variables and set the name of the relationship type as the 1$^{st}$ variable.*

    Supervisor[EmpNo, ProNo(Supervise , , , )]

4. *Chose the remaining 3 variables from the values of the cardinality ratio pairs associated with the relationship type as follows*
    a. *As the 2$^{nd}$ variable, chose the min value of the cardinality ratio pair nearest to the entity type at the N-side of the relationship type,*

    Supervisor[EmpNo, ProNo(Supervise,1 , , )]

    b. *As the 3$^{rd}$ and 4$^{th}$ variables, choose the min & max values of the cardinality ratio pair farthest to the entity type lies at the N-side of the relationship type,*

    Supervisor[EmpNo, ProNo(Supervise, 1 ,1 ,2 )]

5. *Include any set of simple attributes, if existed, of the relationship type as attributes of the chosen relation following the FK included*

    Supervisor[EmpNo, ProNo(Supervise, 1 , 1, 2 ), StartDate]

Following is the completed RDS of the ER model in Figure 3 above.

   Employee[EmpNo, Name, Address, Salary]
   Project[ProNo, Name, Location ]

Supervisor[<u>EmpNo</u>, <u>ProNo</u>(Supervise, 1 , 1, 2 ), StartDate]

**Step SMG: To transform a binary M:N relationship type between a subtype and a regular entity type**

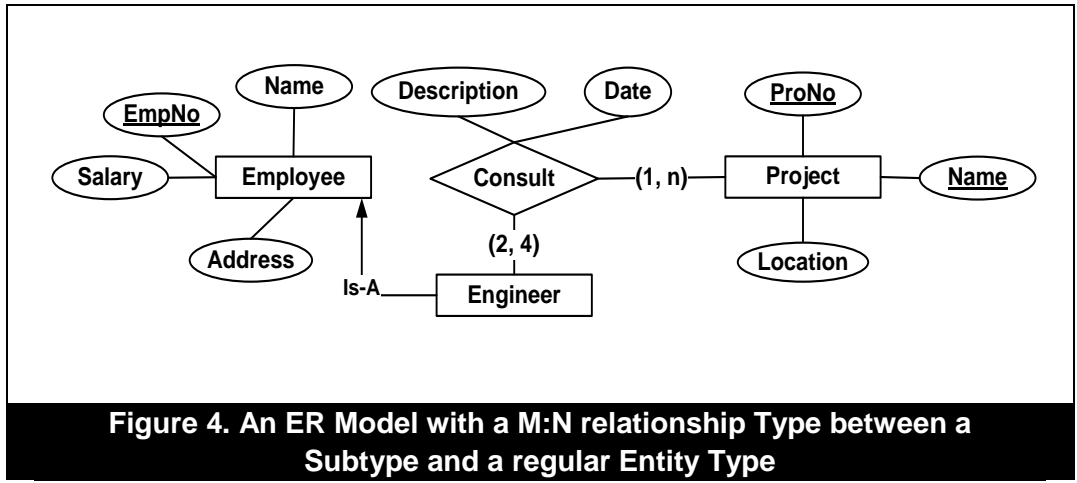

**Figure 4. An ER Model with a M:N relationship Type between a Subtype and a regular Entity Type**

Follow the steps below to transform a M:N relationship type that can be existed between a subtype and a regular entity types.

1. *Create a new relation by the name of the relationship type*,
2. *Include in the relation as its first two attributes, the PK of the subtype relation and the PK of the regular entity type relation*.

    Consult[EmpNo, ProNo ]

3. *Give the including PK of the subtype relation, a prefix, the name of the subtype. Underline the first two attributes separately. These attributes will act in combination as the PK of the new relation, and separately as FKs to refer to their respective relations*. (Note that the "tilde" symbol, " ~ ", is used to connect the prefix to the including PK)

Consult[Eengineer~EmpNo, ProNo ]

4. *Give each of the included PK a bracketed suffix that includes the min-max-cardinality ratio corresponding to the participation of the entity type of the respective PK in the relationship type.*

Consult[Engineer~EmpNo(2, 4), ProNo(1, n) ]

5. *Include any set of simple attributes, if existed, of the relationship type as attributes of the new relation following the suffixed FKs included.*

Consult[ManagerEngineer~EmpNo(2, 4), ProNo(1, n), Description, Date ]

Following is the completed RDS of the ER model in Figure 4 above.

Employee[EmpNo, Name, Address, Salary]
Project[ProNo, Name, Location]
Consult[Engineer~EmpNo(2, 4), ProNo(1, n), Description, Date ]

### Step SOS: Transforming a 1:1 relationship type between two subtypes

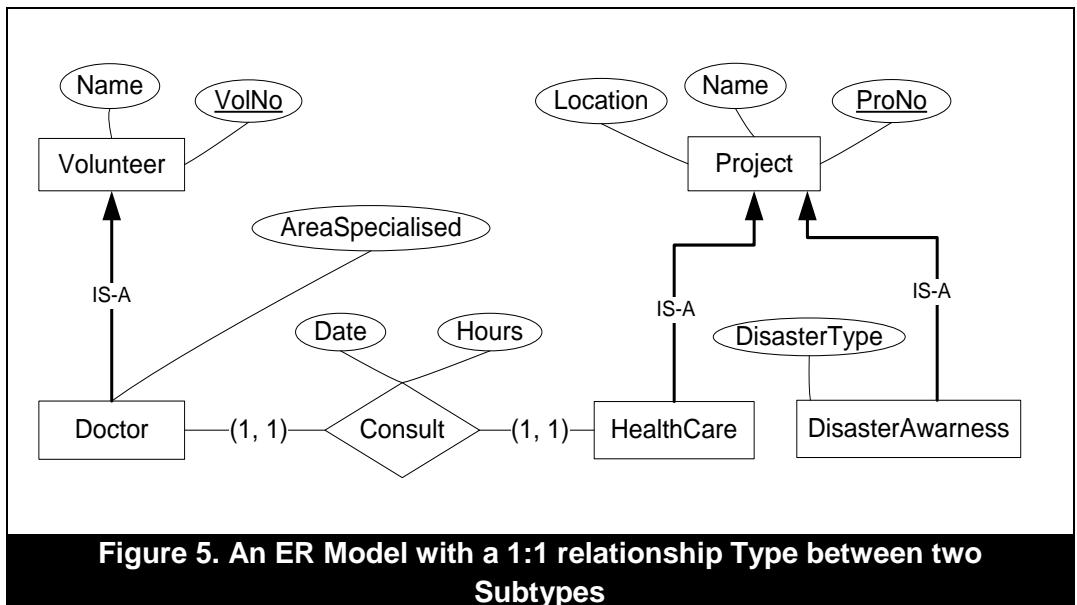

**Figure 5. An ER Model with a 1:1 relationship Type between two Subtypes**

Follow the steps below to transform a 1:1 relationship type that can be existed between two subtypes.

1. *Chose a subtype relation–"Doctor", say–that must have been created following the Step SUB.*

    Doctor[VolNo, AreaSpecialised]

2. *Include as a foreign key (FK) in the chosen relation following its last attribute, the PK of the non-chosen subtype relation. Rename the including PK by giving it a prefix, the name of its corresponding subtype*

    Doctor[VolNo, AreaSpecialised, HealthCare~ProNo]

3. *To the including FK, give a bracketed suffix of 4 variables and set the name of the relationship type as the 1$^{st}$ variable.*

    Doctor[VolNo, AreaSpecialised, HealthCare~ProNo(Consult , , , )]

4. *Chose the remaining 3 variables from the values of the cardinality ratio pairs associated with the relationship type, as follows*
    a. *As the 2$^{nd}$ variable, chose the min value of the cardinality ratio pair nearest to the subtype corresponding to the chosen relation*,

    Doctor[VolNo, AreaSpecialised, HealthCare~ProNo(Consult ,1 , , )]

    b. *As the 3$^{rd}$ and 4$^{th}$ variables, choose the min & max values of the cardinality ratio pair farthest to the subtype corresponding to the chosen relation*,

    Doctor[VolNo, AreaSpecialised, HealthCare~ProNo(Consult ,1 ,1 ,1 )]

5. *Include any set of simple attributes, if existed, of the relationship type as attributes of the chosen relation following the FK included*

    Doctor[VolNo, AreaSpecialised, HealthCare~ProNo(Consult ,1 ,1 ,1 ), Date, Hours]

Following is the completed RDS of the ER model in Figure 5 above.

> Volunteer[<u>VolNo</u>, Name]
>
> Project[<u>ProNo</u>, Name, Location]
>
> Doctor[<u>VolNo</u>, AreaSpecialised, HealthCare~ProNo(Consult ,1 ,1 ,1 ), Date, Hours]

**Step SNS: Transforming a one-to-many relationship type between two subtypes**

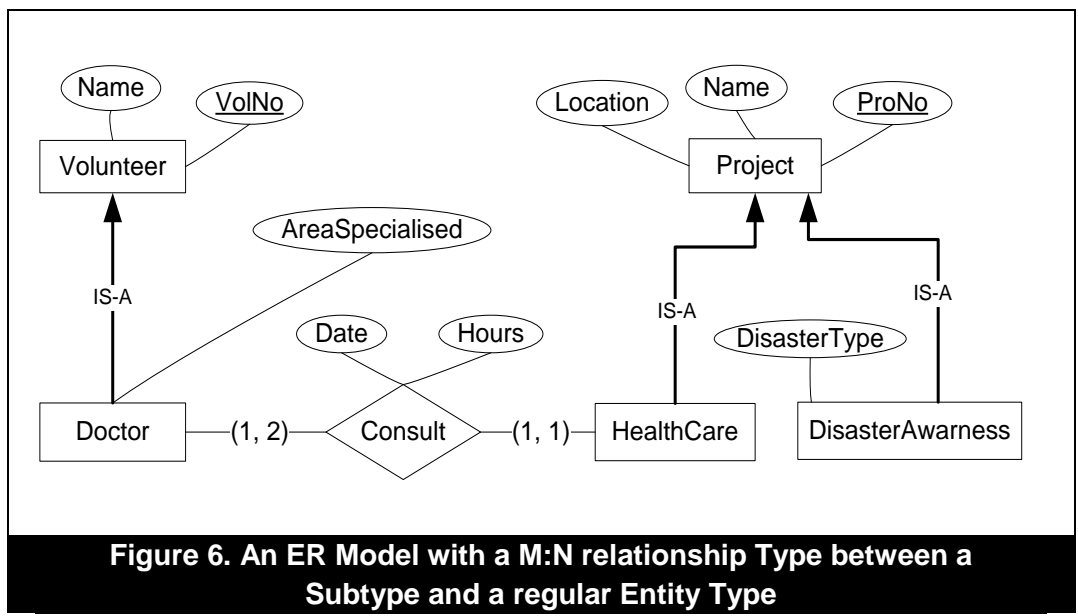

**Figure 6. An ER Model with a M:N relationship Type between a Subtype and a regular Entity Type**

Follow the steps below to transform a 1:N relationship type that can be existed between two subtypes.

1. *Chose the relation analogy to the subtype that lies at the N-side of the relationship type.*

HealthCare[ProNo]

2. Iinclude as a FK in the chosen relation following its last attribute, the PK of the non-chosen subtype relation analogy to the other subtype. Rename the including PK by giving it a prefix, the name of its corresponding subtype

    HealthCare[ProNo, Doctor~VolNo]

6. To the including FK, give a bracketed suffix of 4 variables and chose the name of the relationship type as the $1^{st}$ variable.

    HealthCare[ProNo, Doctor~VolNo(Consult , , , )]

7. Chose the remaining 3 variables from the values of the cardinality ratio pairs associated with the relationship type, as follows
    a. As the $2^{nd}$ variable, chose the min value of the cardinality ratio pair nearest to the entity type at the N-side of the relationship type,

    HealthCare[ProNo, Doctor~VolNo(Consult ,1 , , )]

    b. As the $3^{rd}$ and $4^{th}$ variables, choose the min & max values of the cardinality ratio pair farthest to the entity type at the N-side of the relationship type

    HealthCare[ProNo, Doctor~VolNo(Consult ,1 ,1 ,2) ]

8. Include any set of simple attributes, if existed, of the relationship type as attributes of the chosen relation following the FK included

    HealthCare[ProNo, Doctor~VolNo(Consult ,1 ,1 ,2), Date, Hours ]

Following is the completed RDS of the ER model in Figure 6 above.

   Volunteer[VolNo, Name]
   Project[ProNo, Name, Location]
   DisasterAwareness [ProNo, DisasterType]
   HealthCare[ProNo, Doctor~VolNo(Consult ,1 ,1 ,2), Date, Hours ]

**Step SMS: Transforming a Many-to-many relationship type between two subtypes**

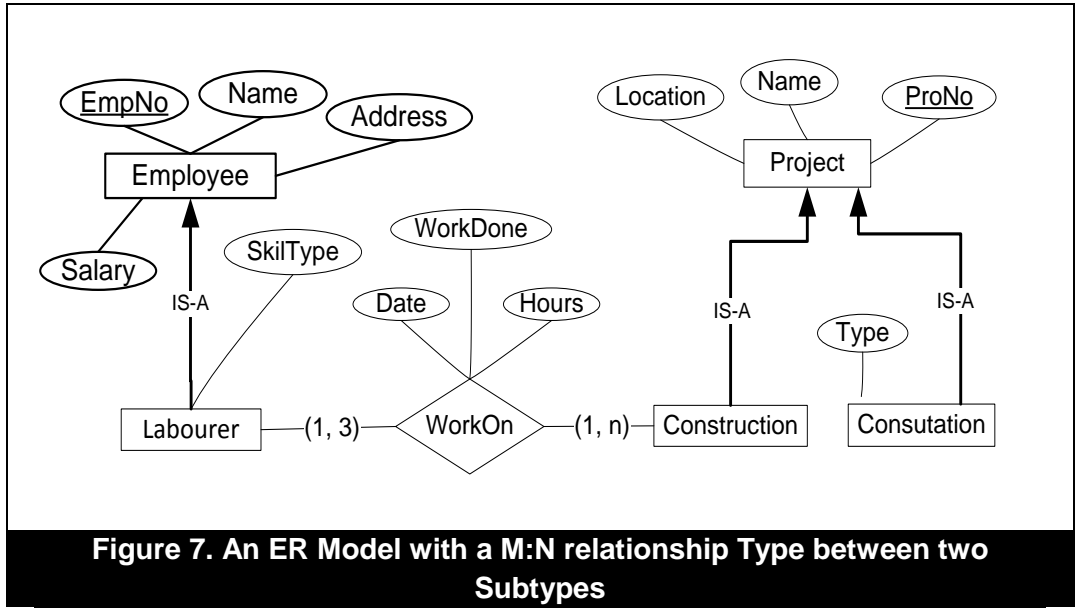

**Figure 7. An ER Model with a M:N relationship Type between two Subtypes**

Follow the steps below to transform a M:N relationship type that can be existed between two subtypes.

1. *Create a new relation by the name of the relationship type*,
2. *Include in the relation as its first two attributes, the PKs of the relations corresponding to the subtypes*.

    WorkOn[EmpNo, ProNo ]

3. *Give each of the including PK, a prefix, the name of the corresponding subtype. Underline the first two attributes separately. These attributes will act in combination as the PK of the new relation, and separately as FKs to refer to their respective relations*.

    Consult[Laborer~EmpNo, Construction~ProNo ]

4. *Suffix each of the included PK, by the min-max-cardinality ratio corresponding to the respective entity types' participation in the relationship*.

Consult[Laborer~EmpNo (1, 3), Construction~ProNo (1, n) ]

5. *Include any set of simple attributes, if existed, of the relationship type as attributes of the new relation following the suffixed FKs included.*

Consult[Laborer~EmpNo (1, 3), Construction~ProNo (1, n), Date, WorkDone, Hours ]

With this transformation the final RDS of the ER model in the Figure 7 will be changed to the following.

Employee[EmpNo, Name, Address, Salary]
Project [ProNo, Name, Location]
Construction [ProNo, Type]
Consult[Laborer~EmpNo (1, 3), Construction~ProNo (1, n), Date, WorkDone, Hours ]

**Step THG: To transform ternary and higher order (n > 3) relationship types among regular entity types**

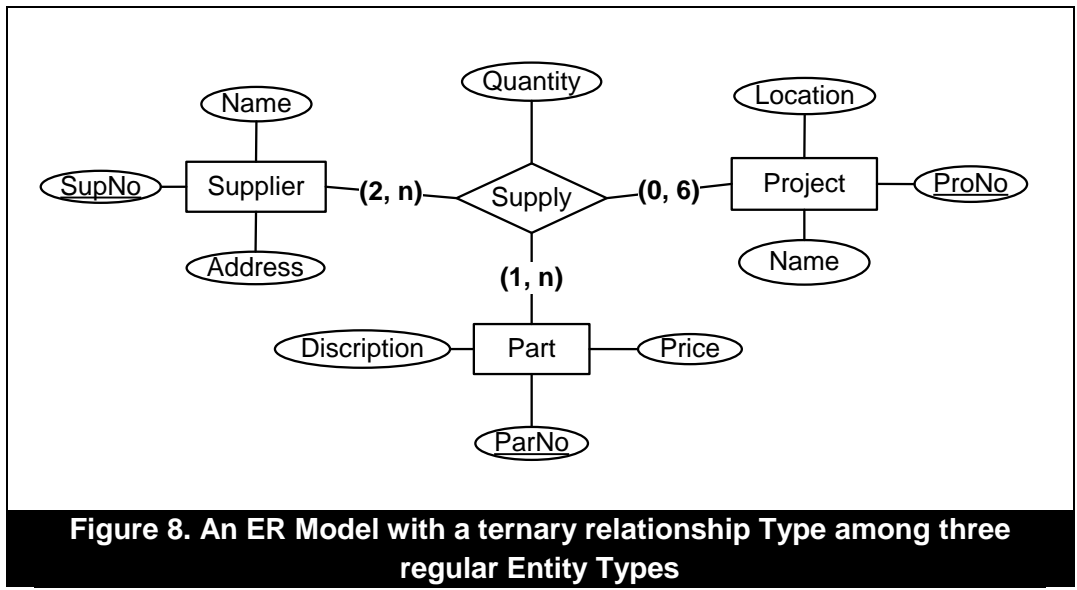

**Figure 8. An ER Model with a ternary relationship Type among three regular Entity Types**

Follow the steps below to transform a ternary relationship type that can be existed between three regular entity types.

1. *Create a new relation by the name of relationship type*,
2. *Include in the new relation as its first three attributes and as FKs, the PKs of the relations corresponding to the regular entity types participated in the relationship type and underline them separately. These attributes will act in combination as the PK of the new relation, and separately as FKs to refer to their respective relations*.

    Supply[SupNo, ParNo, ProNo]

2. *Suffix each of the included FK, with a bracket that includes the min-max-cardinality ratio values respective to the participation in the relationship type of the entity type analogy to the respective FK concerned*,

    Supply[SupNo(2, n), ParNo(1, n), ProNo(0, 6)]

3. *Include any set of simple attributes, if existed, of the relationship type as attributes of the relation following the last suffixed FK included*

    Supply[SupNo(2, n), ParNo(1, n), ProNo(0, 6), Quantity ]

Thus, the final RDS of the ER model in the Figure 8 is as follows.

    Supplier [SupNo, Name, Address]
    Part [ParNo, Description, Price]
    Project [ProNo, Name, Location]
    Supply[SupNo(2, n), ParNo(1, n), ProNo(0, 6), Quantity ]

This is the end of the demonstration of the remained 8 number of steps of the new algorithm and how they can be applied to transform ER models to the relational database model.

*Those who are interested are invited to evaluate the material presented and send feedback to the author.*

Author's contacts: dhammika.pieris@monash.edu, dhammikapieris@yahoo.com